\documentclass[12pt,journal,draftclsnofoot,a4paper,oneside,onecolumn]{IEEEtran}
\usepackage{times,amsmath,epsfig}
\usepackage{cite}
\usepackage{graphicx}
\usepackage{psfrag}
\usepackage{subfigure}
\usepackage{url}
\usepackage{stfloats}
\usepackage{amsmath}
\usepackage{amssymb}
\usepackage{array}

\newcounter{step}
\newlength{\totlinewidth}
  {\end{list}%
  \rule{\linewidth}{1pt}}
\newcounter{substep}

  {\end{list}}

\newlength{\aligntop}
\newlength{\alignbot}
 \makeatletter
\renewenvironment{align}{%
  \vspace{\aligntop}
  \start@align\@ne\st@rredfalse\m@ne
}{%
  \math@cr \black@\totwidth@
  \egroup
  \ifingather@
    \restorealignstate@
    \egroup
    \nonumber
    \ifnum0=`{\fi\iffalse}\fi
  \else
    $$%
  \fi
  \ignorespacesafterend%
  \vspace{\alignbot}\par\noindent
} \makeatother

\title{Differential Modulation for Bi-directional Relaying with Analog Network Coding}

\author{Lingyang~Song, Yonghui~Li, Anpeng Huang,
Bingli~Jiao, and~Athanasios~V.~Vasilakos
\thanks{Copyright (c) 2010 IEEE. Personal use of this material is permitted. However, permission to use this material for any other purposes must be obtained from the IEEE by sending a request to pubs-permissions@ieee.org.}
\thanks{This work was partially supported by the National Natural Science Foundation of China under Grant number 60972009 and 60811130529.}
\thanks{Lingyang~Song, Bingli~Jiao and Anpeng Huang are with Peking University, China (e-mail:
\protect\url{lingyang.song@pku.edu.cn,jiaobl@pku.edu.cn,hapku@pku.edu.cn}).}
\thanks{Yonghui~Li is with University
of Sydney, Australia (e-mail: \protect\url{lyh@ieee.org}).}
\thanks{Athanasios V. Vasilakosis is with University
of Western Macedonia, Greece (e-mail:
\protect\url{vasilako@ath.forthnet.gr}).}}

\begin{document}

\maketitle
\begin{abstract}

In this paper, we propose an analog network coding scheme with
differential modulation~(ANC-DM) using amplify-and-forward protocol
for bidirectional relay networks when neither the source nodes nor
the relay knows the channel state information (CSI). The performance
of the proposed ANC-DM scheme is analyzed and a simple asymptotic
bit error rate (BER) expression is derived. The analytical
results are verified through simulations. It is shown that the BER
performance of the proposed differential scheme is about 3 dB away
from that of the coherent detection scheme. To improve the system
performance, the optimum power allocation between the sources and
the relay is determined based on the simplified BER. Simulation
results indicate that the proposed differential scheme with optimum
power allocation yields 1-2 dB performance improvement over an equal
power allocation scheme.
\end{abstract}

%\pagebreak

%%%%%%%%%%%%%%%%%%%%%%%%%%%%%%%%%%%%%%%%%%%%%%%%%%%%%%%%%%%%%%%%%%%%%%%%%%%%%%%%%%%%%%%%%%%%%%%%%%%
\section{Introduction}
%%%%%%%%%%%%%%%%%%%%%%%%%%%%%%%%%%%%%%%%%%%%%%%%%%%%%%%%%%%%%%%%%%%%%%%%%%%%%%%%%%%%%%%%%%%%%%%%%%%
Bi-directional relay communication has attracted considerable
interest recently
\cite{Ahlswede2000}--\hspace{-.00000765em}\cite{Tao2008}, and
various bi-directional relay protocols for wireless systems have
been proposed
\cite{Ahlswede2000}--\hspace{-.00001mm}\cite{Popovski2007}. In
\cite{Ahlswede2000}~\cite{Yuen2008}, the conventional network coding
scheme is applied to the bi-directional relay network. Two source
nodes transmit to the relay, separately. The relay decodes the
received signals, performs binary network coding, and then
broadcasts network coded symbols back to both source nodes. However,
this scheme may cause irreducible error floor due to the detection
errors which occur at the relay node.

In \cite{Popovski2007}--\hspace{-.00001mm}\cite{Katti2007}, an
amplify and forward based network coding scheme, referred to as the
analog network coding, was proposed. In this scheme, both source
nodes transmit at the same time so that the relay receives a
superimposed signal. The relay then amplifies the received signal
and broadcasts it to both source nodes. Analog network coding is
particularly useful in wireless networks as the wireless channel
acts as a natural implementation of network coding by summing the
wireless signals over the air.

Almost all existing works in bi-directional relay communications
using analog network coding assume that the sources and the
destination have perfect knowledge of channel state
information~(CSI) for all transmission links. As a result, coherent
detection can be readily employed at either sources or relay, or
both \cite{Ahlswede2000}--\hspace{-.00001mm}\cite{Katti2007}. In
some scenarios, e.g. the slow fading environment, the CSI is likely
to be acquired by the use of pilot symbols. However, when the
channel coefficients vary fast, channel estimation may become
difficult. In addition, the channel estimation increases
computational complexity in the relay node and reduces the data
rate. Moreover, it would be difficult for the destination to acquire
the source-to-relay channel perfectly through pilot signal
forwarding without noise amplification. Hence, differential
modulation without need of any CSI would be a practical solution.

In a differential bi-directional relay network, each source receives
a superposition of differentially encoded signals from the other
source, and it has no knowledge of CSI of both channels. All these
problems present a great challenge for designing differential
modulation schemes in two-way relay channels. In \cite{Tao2008},
differential receivers for differential two-way relaying were
presented using analog network coding. These non-coherent schemes
were realized by averaging the channel coefficients of the
substraction of adjacent received signals. However, the approaches
may result in more than 3 dB performance loss compared to the
coherent schemes due to instantaneous detection errors.

In this paper, we propose an analog network coding scheme with
differential modulation~(ANC-DM) using amplify-and-forward protocol
for bidirectional relay networks so that the CSI is not required at
both sources and the relay. The performance of the proposed ANC-DM
scheme is analyzed and a simple asymptotic bit error rate (BER)
expression is derived. The analytical results are verified through
simulations. They show that the proposed differential scheme is
about 3 dB away compared to the coherent detection scheme. To
improve the system performance, the optimum power allocation between
the sources and the relay is determined based on the provided
simplified BER. Simulation results show that optimum power allocation yields 1-2 dB
performance improvement over an equal power allocation
scheme.

Note that unlike \cite{Tao2008}, the destination realizes
differential detection by subtracting away its own contribution in
the received signals based on the power estimation. Besides, our scheme requires linear complexity, while the
detectors in~\cite{Tao2008} are much complex and non-linear which
may calculate the modified Bessel function of second kind. Note also
that the application of differential modulation for the
bi-directional relaying in \cite{Ahlswede2000}\cite{Yuen2008} with
digital network coding is relatively straightforward as the received
signals at the relay from the sources can be decoded separately due
to the use of orthogonal transmissions. However, for
amplify-and-forward relaying
\cite{Popovski2007}--\hspace{-.00001mm}\cite{Katti2007}, the relay
receives a superposition of differentially encoded signals from the
sources which makes the final detection at the source difficult.

The rest of the paper is organized as follows: In Section~II, we
describe the proposed differential modulation scheme. Theoretical
analysis is given in Section~III. Section~IV presents the optimal
transmit power allocation between the sources and the relay.
Simulation results are provided in Section~V. In Section~VI, we draw
the main conclusions.

\emph{\textbf{Notation}}: Boldface lower-case letters denote
vectors, $(\cdot)^{*}$ stands for complex conjugate, $(\cdot)^{T}$
represents transpose, $\mathbb{E}$ is used for expectation,
$\texttt{Var}$ represents variance,
$\|\textbf{x}\|^2=\textbf{x}^H\textbf{x}$, and $\mathfrak{R}(\cdot)$
denotes real part.
%%%%%%%%%%%%%%%%%%%%%%%%%%%%%%%%%%%%%%%%%%%%%%%%%%%%%%%%%%%%%%%%%%%
\section{Differential Modulation for bidirectional relay networks}
%%%%%%%%%%%%%%%%%%%%%%%%%%%%%%%%%%%%%%%%%%%%%%%%%%%%%%%%%%%%%%%%%%%
%%%%%%%%%%%%%%%%%%%%%%%%%%%%%%%%%%%%%%%%%%%%%%%%%%%%%%%%%%%%%%%%%%%
\subsection{Differential Encoding}
\begin{figure}[]
\centering
\includegraphics[height=5.0in,width=6.6in]{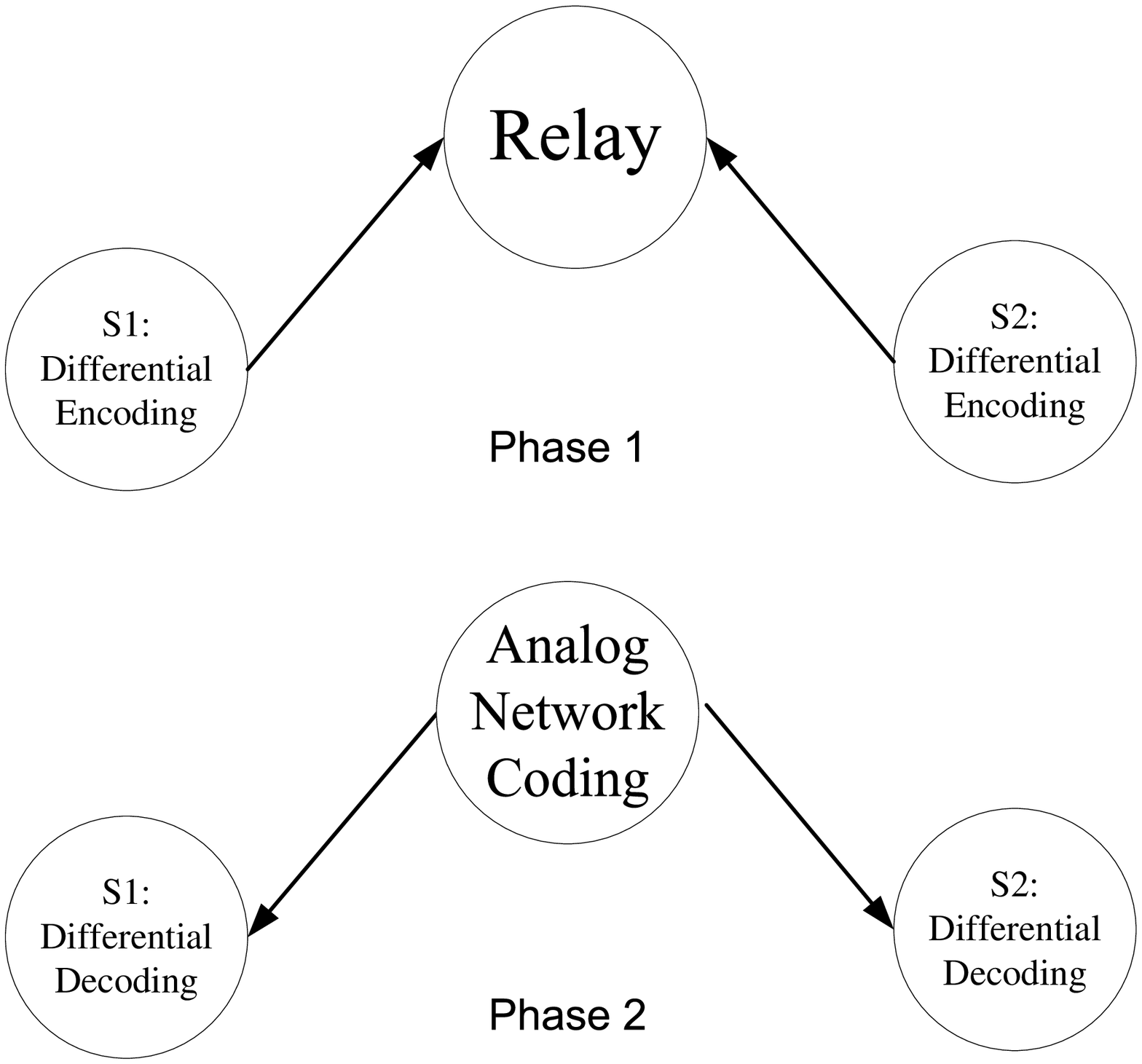}
\caption{Block diagram of the proposed ANC-DM scheme.}
\label{Fig:Diff_bidirectional_relay_dia}
\end{figure}

%%%%%%%%%%%%%%%%%%%%%%%%%%%%%%%%%%%%%%%%%%%%%%%%%%%%%%%%%%%%%%%%%%%
We consider a three-node bi-directional relay network consisting of
two source nodes, denoted by $S_1$ and $S_2$, and one relay node,
denoted by $R$. All nodes are equipped with one antenna and operate
in a half-duplex way so that the complete transmission can be
divided into two phases, as shown in
Fig.~\ref{Fig:Diff_bidirectional_relay_dia}. In the first phase,
both source nodes simultaneously send the differentially encoded
information to the relay, and in the second phase, the relay
broadcasts the combined signals to both sources. Let
$c_1(t){\in}\mathcal{A}$ denote the symbol to be transmitted by the
source $S_1$ at the time $t$, where $\mathcal{A}$ represents a unity
power $M$-PSK constellation set. In the differential modulation
bi-directional relay system, the signal $s_1(t)$ sent by the source
$S_1$ is given by
%%--
\begin{equation}
    s_1(t)=s_1(t-1){c_1(t)}, \hspace{1em}c_1(t){\in}\mathcal{A}
 \label{Eq:s1}
\end{equation}
%%--
Similarly, the signal transmitted by $S_2$ at the time $t$ is given
by
%%--
\begin{equation}
    s_2(t)=s_2(t-1){c_2(t)}, \hspace{1em}c_2(t){\in}\mathcal{A}
 \label{Eq:s2}
\end{equation}
%%--
%%%%%%%%%%%%%%%%%%%%%%%%%%%%%%%%%%%%%%%%%%%%%%%%%%%%%%%%%%%%%%%%%%%
\subsection{Differential Decoding}
%%%%%%%%%%%%%%%%%%%%%%%%%%%%%%%%%%%%%%%%%%%%%%%%%%%%%%%%%%%%%%%%%%%
In the bi-directional relayed transmission, the source nodes first
broadcast the information to the relay. For simplicity, we assume
that the fading coefficients are constant over one frame of length $L$, and change
independently from one frame to another. The received signal in the
relay at time $t$ can be expressed as
%%--
\begin{equation}
    y_r(t)=\sqrt{p_{1}}h_1s_1(t)+\sqrt{p_{2}}h_2s_2(t)+n_r(t).
 \label{Eq:yr}
\end{equation}
%%--
where $p_{1}$ and $p_{2}$ represent the transmit power at $S_1$ and
$S_2$ respectively, $h_1$ and $h_2$ are the Rayleigh fading
coefficients with zero mean and unit variance between $S_1$ and $R$,
and between $S_2$ and $R$, respectively, $n_r(t)$ denotes zero mean
complex Gaussian random variable with two sided power spectral
density of $N_0/2$ per dimension, and we furthermore assume $S_1$,
$S_2$, and $R$ have the same noise variance.

In the second phase, the relay $R$ amplifies $y_r(t)$ by a factor
$\beta$ and then broadcasts its conjugate, denoted by $y_r^*(t)$, to
both $S_1$ and $S_2$. The corresponding signal
 received by $S_1$ at time $t$, denoted by $y_1(t)$, can be
written as
%%--
\begin{align}
    y_1(t)&={\beta}\sqrt{p_{r}}h_1y_r^*(t)+n_1(t)
    \nonumber \\
          &={\mu}s_1^*(t)+{\nu}s_2^*(t)+w_1(t),
 \label{Eq:y1}
\end{align}
%%--
where $\beta=(p_{1}|h_1|^2+p_{2}|h_2|^2+N_0)^{-\frac{1}{2}}$,
$p_{r}$ represents the transmit power by the relay,
$\mu\triangleq{\beta}\sqrt{p_{1}p_r}|h_1|^2$,
$\nu\triangleq{\beta}\sqrt{p_{2}p_r}h_1h_2^*$, and
$w_1(t)\triangleq{\beta}\sqrt{p_r}h_1n_r^*(t)+n_1(t)$. Note that
unlike traditional ANC schemes
\cite{Popovski2007}--\hspace{-.00001mm}\cite{Katti2007}, $y_r^*(t)$
is transmitted from the relay, which obviously yields $\mu>0$. The
reason of doing this is to make the receiver easily estimate $\mu$
and then subtract ${\mu}s_1^*(t)$ in (\ref{Eq:y1}) for differential
detection.

As the relay has no CSI, the normalization factor $\beta$ has to be
obtained indirectly. We may rewrite the received signals in
(\ref{Eq:yr}) in a vector format, given by
%%--
\begin{align}
    \textbf{y}_r=\sqrt{p_{1}}h_1\textbf{s}_1+\sqrt{p_{2}}h_2\textbf{s}_2+\textbf{n}_r,
 \label{Eq:yrvec}
\end{align}
%%--
where $\textbf{y}_r=[{y}_r(1),\ldots,{y}_r(L)]^T$,
$\textbf{s}_1=[{s}_1(1),\ldots,{s}_1(L)]^T$,
$\textbf{s}_2=[{s}_2(1),\ldots,{s}_2(L)]^T$, and
$\textbf{n}_r=[{n}_r(1),\ldots,{n}_r(L)]^T$. To estimate the average receive power, we multiply the
received signals by its Hermitian transpose as
%%--
\begin{align}
    \|\textbf{y}_r\|^2=p_{1}|h_1|^2\textbf{s}_1^H\textbf{s}_1
            &+p_{2}|h_2|^2\textbf{s}_2^H\textbf{s}_2
            +2\sqrt{p_1p_2}\mathfrak{R}\{h_1^*h_2\textbf{s}_1^H\textbf{s}_2\}
            \nonumber \\
            &+2\sqrt{p_1}\mathfrak{R}\{h_1\textbf{n}_r^H\textbf{s}_1\}
            +2\sqrt{p_2}\mathfrak{R}\{h_2\textbf{n}_r^H\textbf{s}_2\}
            +\textbf{n}_r^H\textbf{n}_r
 \label{Eq:betacorr}
\end{align}
%%--
By taking the expectation of (\ref{Eq:betacorr}), $\beta$ can be
then approximated at high SNR by
%%--
\begin{align}
    \beta=\sqrt{\frac{\mathbb{E}\{\textbf{y}_r^H\textbf{y}_r\}}{L}}\approx\sqrt{\frac{\|\textbf{y}_r\|^2}{L}},
 \label{Eq:beta}
\end{align}
%%--
where
$\mathbb{E}\{\textbf{s}_1^H\textbf{s}_1\}=\mathbb{E}\{\textbf{s}_2^H\textbf{s}_2\}=L$,
$\mathbb{E}\{\textbf{n}_r^H\textbf{n}_r\}=LN_0$, and
$\mathbb{E}\{\textbf{s}_1^H\textbf{s}_2\}=\mathbb{E}\{\textbf{n}_r^H\textbf{s}_1\}=\mathbb{E}\{\textbf{n}_r^H\textbf{s}_2\}=0$.

Similarly, the received signal at $S_2$ can be calculated as
%%--
\begin{equation}
    y_2(t)={\beta}\sqrt{p_r}h_2y_r^*(t)+n_2(t),
    %\nonumber \\
          %&={\beta}|h_2|^2s_2(t)+{\beta}h_1^*h_2s_1(t)^*+{\beta}h_2n_r(t)^*+n_2(t),
 \label{Eq:y2}
\end{equation}
%%--
As $S_1$ and $S_2$ are mathematically symmetrical, as shown in
(\ref{Eq:y1}) and (\ref{Eq:y2}), for simplicity, we in the next only
discuss the decoding as well as the corresponding theoretical
analysis for signals received by $S_1$.

Recalling the differential encoding process in (\ref{Eq:s2}),
(\ref{Eq:y1}) can be further written as
%%--
\begin{align}
    y_1(t)&={\mu}s_1^*(t)+{\nu}s_2^*(t)+w_1(t),
        \nonumber \\
          &={\mu}s_1^*(t)+{\nu}s_2^*(t-1)c_2^*(t)+w_1(t),
 \label{Eq:y1d}
\end{align}
%%--
Obviously, since $s_1(t)$ is known, to decode $c_2(t)$ in
(\ref{Eq:y1d}), ${\mu}s_1^*(t)$ should be removed. To do this, we
have to first estimate $\mu$, where $\mu>{0}$.

\begin{figure}[]
\centering
\includegraphics[height=4.5in,width=6.0in]{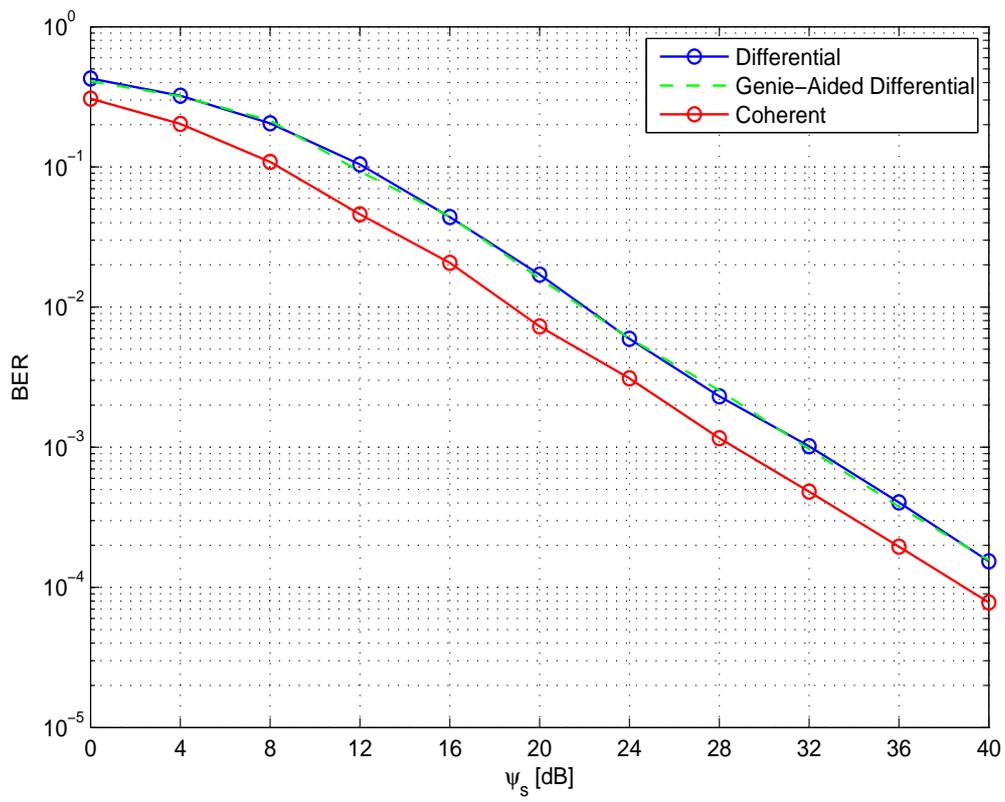}
\caption{Simulated BER performance by differential and coherent
detections, where $p_s=p_r=p/3$.}
\label{Fig:Diff_bidirectional_relay}
\end{figure}

Similar to (\ref{Eq:yrvec}), the received signal in (\ref{Eq:y1})
can be expressed in a vector form as follows
%%--
\begin{align}
    \textbf{y}_1={\mu}\textbf{s}_1+{\nu}\textbf{s}_2+\textbf{w}_1,
 \label{Eq:yrvec}
\end{align}
%%--
where $\textbf{y}_1=[{y}_1(1),\ldots,{y}_1(L)]^T$ and
$\textbf{w}_1=[{w}_1(1),\ldots,{w}_1(L)]^T$. At high SNR, we may
approximately obtain
%%--
\begin{equation}
    \mu^2+|\nu|^2\approx{\textbf{y}_1^H\textbf{y}_1}/L.
 \label{Eq:uv}
\end{equation}
%%--
Since the source node $S_1$ can retrieve its own information
$s_1(t-1)$ and $c_1(t)$, based on (\ref{Eq:s1}) and (\ref{Eq:y1d}),
we have the following transformation
%%--
\begin{align}
    \widetilde{y}_1(t)&\triangleq{c_1^*(t)}y_1(t-1)-y_1(t)
        \nonumber \\
          &={\nu}s_2^*(t-1)\left(c_1(t)-c_2(t)\right)^*
    +\widetilde{w}_1(t),
 \label{Eq:y1y0}
\end{align}
%%--
where $\widetilde{w}_1(t)\triangleq{{c_1^*(t)}w_1(t-1)+w_1(t)}$. Then,
$|\nu|^2$ can be approximately calculated in a similar way as~
(\ref{Eq:uv})
%%--
\begin{align}
    |\nu|^2\approx\frac{\widetilde{\textbf{y}}_1^H\widetilde{\textbf{y}}_1}{L\mathbb{E}\left[|s_2(t-1)|^2\right]\mathbb{E}\left[|c_1(t)-c_2(t)|^2\right]},
 \label{Eq:nu}
\end{align}
%%--
where
$\widetilde{\textbf{y}}_1=[\widetilde{y}_1(1),\ldots,\widetilde{y}_1(L-1)]^T$,
$\mathbb{E}[|s_2(t-1)|^2]=1$ , and the calculation of
$\mathbb{E}[|c_1(t)-c_2(t)|^2]$ is given in Section-II-C. As $\mu$
is positive, by combining (\ref{Eq:beta}), (\ref{Eq:uv}) and
(\ref{Eq:nu}), we have
%%--
\begin{align}
    \mu\approx\left\{
    \begin{array}{ll}
        \sqrt{\Delta}, &\Delta>{0}\\
        0,             &\Delta\leq{0}
    \end{array}
       \right.
 \label{Eq:muf}
\end{align}
%%--
where $\Delta\triangleq
{\frac{{\textbf{y}_1^H\textbf{y}_1}}{L}-\frac{\widetilde{\textbf{y}}_1^H\widetilde{\textbf{y}}_1}{L\mathbb{E}\left[|s_2(t-1)|^2|c_1(t)-c_2(t)|^2\right]}}$.
Note that in low SNR, $\Delta$ could be negative due to the noise
effect, and thus we set $\mu\approx{0}$ instead. The estimation method given in (\ref{Eq:muf}) is evaluated
in Fig.~\ref{Fig:Diff_bidirectional_relay}.

By subtracting ${\mu}s_1^*(t)$, (\ref{Eq:y1d}) can be further
written as
%%--
\begin{align}
    y'_1(t)&\triangleq{y_1(t)}-\mu{s_1(t)}
    \nonumber \\
    &={\nu}s_2^*(t-1)c_2^*(t)+w_1(t)
    \nonumber \\
    &=\left(y'_1(t-1)-w_1(t-1)\right)c_2^*(t)+w_1(t),
 \label{Eq:y1d1}
\end{align}
%%--
Finally, the following linear decoder can be used to recover $c_2(t)$
%%--
\begin{align}
    \widetilde{c}_2&(t)=\text{arg}\underset{c_2(t)\in{\mathcal{A}}}{\max}\text{Re}\left\{y'_1(t){y'_1}^*(t-1)c_2(t)\right\}.
 \label{Eq:mldafk}
\end{align}
And $c_1(t)$ can be differentially decoded in a similar way by the
source $S_2$. Note that in comparison to traditional differential
modulation \cite{Proakis-Digital-Comms}, the extra complexity comes
from the $\mu$ estimation in~(\ref{Eq:muf}), which is linear and
only comprises a few number of additions and multiplications. But
the receiver in~\cite{Tao2008} requires complicated computations,
such as the zeroth-order modified Bessel function of the second
kind.

By ignoring the second order term, the corresponding SNR of the
proposed differential detection scheme can be
written as
%%--
\begin{align}
    \gamma_d&\approx\frac{|{\nu}|^2}{2\texttt{Var}\{w_1(t)\}}
    \nonumber \\
    &\approx\frac{{\beta^2}p_2p_r|h_1|^2|h_2|^2}{2\left({\beta^2}p_rN_0|h_1|^2+N_0\right)}
    \nonumber \\
    &\approx\frac{\psi_2\psi_r|h_1|^2|h_2|^2}{2\left((\psi_1+\psi_r)|h_1|^2+\psi_2|h_2|^2+1\right)},
 \label{Eq:SNRy1}
\end{align}
%%--
where $\texttt{Var}\{w_1(t)\}={\beta^2}p_rN_0|h_1|^2+N_0$,
$\psi_1\triangleq{p_1/N_0}$, $\psi_2\triangleq{p_2/N_0}$, and
$\psi_r\triangleq{p_r/N_0}$.

Alternatively, if coherent detection is used, under the assumption
of $h_1$, $h_2$ and $N_0$ available at $S_1$, after subtracting
${\mu}s_1(t)^*$ from $y_1(t)$ in (\ref{Eq:y1}), the corresponding
SNR can be calculated as
%%--
\begin{align}
    \gamma_c&\triangleq\frac{|{\nu}|^2}{\texttt{Var}\{w_1(t)\}}
    \nonumber \\
    &=
    \frac{\psi_2\psi_r|h_1|^2|h_2|^2}{(\psi_1+\psi_r)|h_1|^2+\psi_2|h_2|^2+1}.
 \label{Eq:coherent}
\end{align}
%%--

By comparing (\ref{Eq:SNRy1}) and (\ref{Eq:coherent}), we can easily
obtain
%%--
\begin{align}
    \gamma_d\approx{\frac{\gamma_c}{2}}
 \label{Eq:cod}
\end{align}
%%--
which clearly indicates that the differential detector in
(\ref{Eq:mldafk}) suffers around 3 dB performance loss compared to
the coherent scheme.
%%%%%%%%%%%%%%%%%%%%%%%%%%%%%%%%%%%%%%%%%%%%%%%%%%%%%%%%%%%%%%%%%%%%%%%%%%%%%%%%%%%%%%%%%%%%%%%%%%%
\subsection{The Calculation of $\mathbb{E}[|c_1(t)-c_2(t)|^2]$ in (\ref{Eq:nu})}
%%%%%%%%%%%%%%%%%%%%%%%%%%%%%%%%%%%%%%%%%%%%%%%%%%%%%%%%%%%%%%%%%%%%%%%%%%%%%%%%%%%%%%%%%%%%%%%%%%%
From (\ref{Eq:nu}), the average power of
$c_1(t)-c_2(t)$ needs to be calculated. When $M$-PSK constellations
are applied, the number of symbols produced in the new constellation
by $c_1(t)-c_2(t)$ is finite. Hence, it is easy to derive the
average power of the new constellation sets. Note that the value of
$c_1(t)-c_2(t)$ can be equal to zero, which may affect the
estimation accuracy in (\ref{Eq:nu}).

In order to overcome this problem, we may properly choose a rotation
angle for the symbol modulated in source $S_2$ by
$c_2(t)e^{-j\theta}$, ensuring that $c_1(t)-c_2(t)$ in (\ref{Eq:nu})
is nonzero. For a $M$-PSK constellation, the effective rotation
angle is in the interval $[-\pi/M,\pi/M]$ from the symmetry of
symbols. For a regular and symmetrical constellation, the rotation
angle may be simply set as $\theta=\pi/M$. Similar approach may be
used to generate the rotation angle for other types of
constellations.

Here, we give two examples on how to compute the average symbol power:

1. Supposing BPSK constellation $\{-1,1\}$ is used, we have $c_1(t) - c_2(t)\in \{-2, 0, 2\}$. Hence, the calculation of average power in the new set is straightforward.

2. Supposing $S_1$ uses the BPSK set $\{-1,1\}$, by constellation rotation, $S_2$ can use $\{-j,j\}$. And we can get $c_1(t) - c_2(t)\in  \{-1-j, -1+j, 1-j, 1+j\}$. Hence, it is also easy to derive the average power in the new constellation set.

%%%%%%%%%%%%%%%%%%%%%%%%%%%%%%%%%%%%%%%%%%%%%%%%%%%%%%%%%%%%%%%
\section{Performance Analysis}
%%%%%%%%%%%%%%%%%%%%%%%%%%%%%%%%%%%%%%%%%%%%%%%%%%%%%%%%%%%%%%%
For simplicity, we in this section analyze the BER performance using BPSK for the proposed ANC-DM scheme, and we assume $p_1=p_2=p_s$ and $p_s=\lambda{p}_r$,
where $\lambda>{0}$, and thus, $\psi_1=\psi_2=\psi_s=\lambda\psi_r$. (\ref{Eq:SNRy1}) can be rewritten as
%%--
\begin{align}
    \gamma_d\approx\frac{\psi_s\psi_r'|h_1|^2|h_2|^2}{2(1+\lambda)\left(\psi_r'|h_1|^2+\psi_s|h_2|^2+1\right)},
 \label{Eq:SNRy11}
\end{align}
%%--
where $\psi_r'\triangleq(1+\lambda)\psi_r$.

Let $X=\gamma_d$, and the BER for BPSK modulation can generally be expressed as
%%--
\begin{align}
    \text{BER}=\mathbb{E}\left[Q(2X)\right]=\frac{1}{2\sqrt{\pi}}\int_0^\infty{\frac{\exp(-x)F_X(x)}{\sqrt{x}}\text{d}x},
 \label{Eq:aveSER}
\end{align}
%%--
where $Q(\cdot)$ is the Gaussian-Q function, $F_X(x)$ is the cumulative distribution function (CDF) of $X$. The right side of the equation can be readily obtained by integration by parts. The above expression is useful as it allows us to obtain the BER directly in terms of the CDF of $X$.

By using a general result from \cite{Wang2003}, the BER in (\ref{Eq:aveSER}) can be approximated in the high SNR regime by considering a first order expansion of the CDF of $X$. Specifically, if the first order expansion of the CDF of $X$ can be written in the form
\begin{equation}
    F_X(x)
    =
    \frac{\alpha{x^{N+1}}}{\overline{\lambda}^{N+1}(N+1)}+o(x^{N+1+\varepsilon}),\varepsilon>0,
 \label{Eq:CDFexpansion}
\end{equation}
%%--
where $\overline{\lambda}$ represents the average transmit SNR. At high SNR, the asymptotic BER is given by \cite{Wang2003}
%%--
\begin{align}
    \text{BER}=\frac{\alpha{\Gamma(N+\frac{3}{2})}}{2\sqrt{\pi}\overline{\lambda}^{N+1}(N+1)}+o\left((\overline{\lambda})^{-(N+1)}\right).
 \label{Eq:aveSERaym}
\end{align}
%%--

The PDF of $X$ can be obtained with the help
of~\cite{Hasna}
%%--
\begin{align}
    P_{X}(x)=
    \frac{8(1+\lambda)^2x\exp\left(-2(1+\lambda)(\psi_{s}^{-1}+\psi_r'^{-1})x\right)}{\psi_{s}\psi_r'}
    &\left[\frac{\psi_{s}+\psi_r'}{\sqrt{\psi_{s}\psi_r'}}
            \right.
    {\times}K_1\left(\frac{4(1+\lambda)x}{\sqrt{\psi_{s}\psi_r'}}\right)
    \nonumber \\
    &\vspace{-3em}\left.+2K_0\left(\frac{4(1+\lambda)x}{\sqrt{\psi_{s}\psi_r'}}\right)\right]U\left(2(1+\lambda)x\right),
 \label{Eq:PDFaafk}
\end{align}
%%--
where $K_0(\cdot)$ and $K_1(\cdot)$ are the zeroth-order and first-order modified Bessel functions of the second kind,
respectively, and $U(\cdot)$ is the unit step function. Note that the exact BER, which is complicated in computation, does
not have a closed-form solution. However, at high SNR, when $z$
approaches zeros, the $K_1(z)$ function converges to $1/z$
\cite{Abramowitz}, and the value of the $K_0(z)$ function is
comparatively small, which could be ignored for asymptotic analysis.
Hence, $P_{X}(x)$ in (\ref{Eq:PDFaafk}) can be approximated as
%%--
\begin{align}
    P_{X}(x)\approx
    \frac{2(1+\lambda)(\psi_{s}+\psi_r')\exp\left(-2(1+\lambda)(\psi_{s}^{-1}+\psi_r'^{-1})x\right)}{\psi_{s}\psi_r'}.
 \label{Eq:PDFPx}
\end{align}
%%--

For the ANC-DM, the CDF of destination SNR $\gamma_d$ can be approximated as
\begin{align}
    F_X(x)
    &\approx
    1-\exp\left(-2(1+\lambda)(\psi_{s}^{-1}+\psi_r'^{-1})x\right)
    \nonumber \\
    &\approx
    2(1+\lambda)(\psi_{s}^{-1}+\psi_r'^{-1})x+o(x^{1+\varepsilon}).
 \label{Eq:CDapprox}
\end{align}
%%--
Finally, comparing (\ref{Eq:CDapprox}) with (\ref{Eq:CDFexpansion}) and (\ref{Eq:aveSERaym}), the asymptotic BER of ANC-DM at high SNR can be approximated as
%%--
\begin{align}
    \text{BER}&\approx\frac{2(1+\lambda){\Gamma(\frac{3}{2})}}{2\sqrt{\pi}}(\psi_{s}^{-1}+\psi_{r}'^{-1})
    \nonumber \\
    &=\frac{(1+\lambda)(\psi_{s}^{-1}+\psi_{r}'^{-1})}{2}.
 \label{Eq:upPEPrs}
\end{align}

%%%%%%%%%%%%%%%%%%%%%%%%%%%%%%%%%%%%%%%%%%%%%%%%%%%%%%%%%%%%%%
\section{Transmit Power Allocation}
%%%%%%%%%%%%%%%%%%%%%%%%%%%%%%%%%%%%%%%%%%%%%%%%%%%%%%%%%%%%%%%
In this section, we discuss how to allocate power to both sources
and the relay subject to total transmission power constraint. It can
be seen from (\ref{Eq:upPEPrs}) that the asymptotic BER of the
proposed differential modulation scheme depends non-linearly upon
$p_{s}$ and $p_{r}$. Hence, when the total transmit power is fixed,
$2p_{s}+p_{r}=p$, the power allocation problem over Rayleigh
channels can be formulated to minimize the asymptotic BER at high SNR
in~(\ref{Eq:upPEPrs})
%%--
\begin{align}
    &\min \text{BER}
    \nonumber \\
    &\text{s.t.} \hspace{1mm} 2p_{s}+p_{r}=p\hspace{3mm} (0{<}p_{s}{<}p,\hspace{1mm} 0{<}p_{r}{<}p),
 \label{Eq:RSpower}
\end{align}
%%--
where we assume $p_1=p_2=p_s$ and $p_s=\lambda{p_r}$.

The power allocation problem is to find $p_s$ such that the BER in
(\ref{Eq:upPEPrs}) is minimized subject to the power constraint by
solving the following optimization problem
%%--
\begin{equation}
    \mathcal{L}(p_s)=\text{BER}+\xi(2p_{s}+p_{r}-p),
 \label{Eq:FLag}
\end{equation}
%%--
where $\xi$ is a positive Lagrange multiplier. The necessary
condition for the optimality is found by setting the derivatives of
the Lagrangian in (\ref{Eq:FLag}) with respect to $p_{s}$ and
$p_{r}$ equal to zero, respectively. Reusing the power constraint,
we can calculate at high SNR that
%%--
\begin{align}
    &p_s=\frac{p}{4},
    \nonumber \\
    &p_r=\frac{p}{2},
 \label{Eq:PowerAll}
\end{align}
%%--
which indicates the power allocated in the relay should be equal to
the total transmit power at both sources in order to compensate the
energy used to broadcast combined information in one time slot.
%%%%%%%%%%%%%%%%%%%%%%%%%%%%%%%%%%%%%%%%%%%%%%%%%%%%%%%%%%%%%%%
\section{Simulation Results}
%%%%%%%%%%%%%%%%%%%%%%%%%%%%%%%%%%%%%%%%%%%%%%%%%%%%%%%%%%%%%%%
In this section, we provide simulation results for the proposed
ANC-DM scheme. We also include corresponding coherent detection
results for comparison. All simulations are performed for a BPSK
modulation over the Rayleigh fading channels. The frame length is
$L=100$. For simplicity, we assume that $2p_s+p_r=p=3$, and $S_1$, $S_2$ and $R$ have
the same noise variance $N_0$. The SNR $\psi_s$ can be then calculated as $\psi_s=p_s/N_0$.

Fig.~\ref{Fig:Diff_bidirectional_relay} shows the simulated
BER performance for differential and coherent ANC schemes in
bi-directional relaying without using constellation rotation. Equal transmit power allocation is applied:
$p_s=p_r=p/3$. It can be seen that the differential scheme suffers
around 3 dB performance loss compared to the coherent ANC scheme,
which has been validated by (\ref{Eq:cod}). We also include the
Genie-aided result by assuming that $\mu$ is perfectly known by the
source such that traditional differential decoding can be performed.
It shows from the results that there is almost no performance loss
using the estimation method in (\ref{Eq:muf}) which clearly
justifies the robustness of the proposed differential decoder.

It is worthwhile mentioning that, in \cite{Tao2008}, it uses similar
Genie-aided result as a benchmark as well. The major difference is
that the detectors in \cite{Tao2008} have much inferior performance
than the Genie-aided result, and it has about 6dB performance loss in comparison to the coherent detection scheme. However, our proposed detection algorithm
has comparable performance with the genie-aided result, and only has 3dB performance loss than the coherent detection results. This clearly indicates that our proposed
method outperforms the differential detectors in~\cite{Tao2008}. The main performance loss in \cite{Tao2008} is due to that uncoherent detection approach is employed by statistically averaging off the impact of channel fading coefficients ignoring the instantaneous channel state information. But, our method utilizes differential detection relying on the operation with the previous received signals which could be more adaptive to variation of the channels.

\begin{figure}[]
\centering
\includegraphics[height=4.5in,width=6.0in]{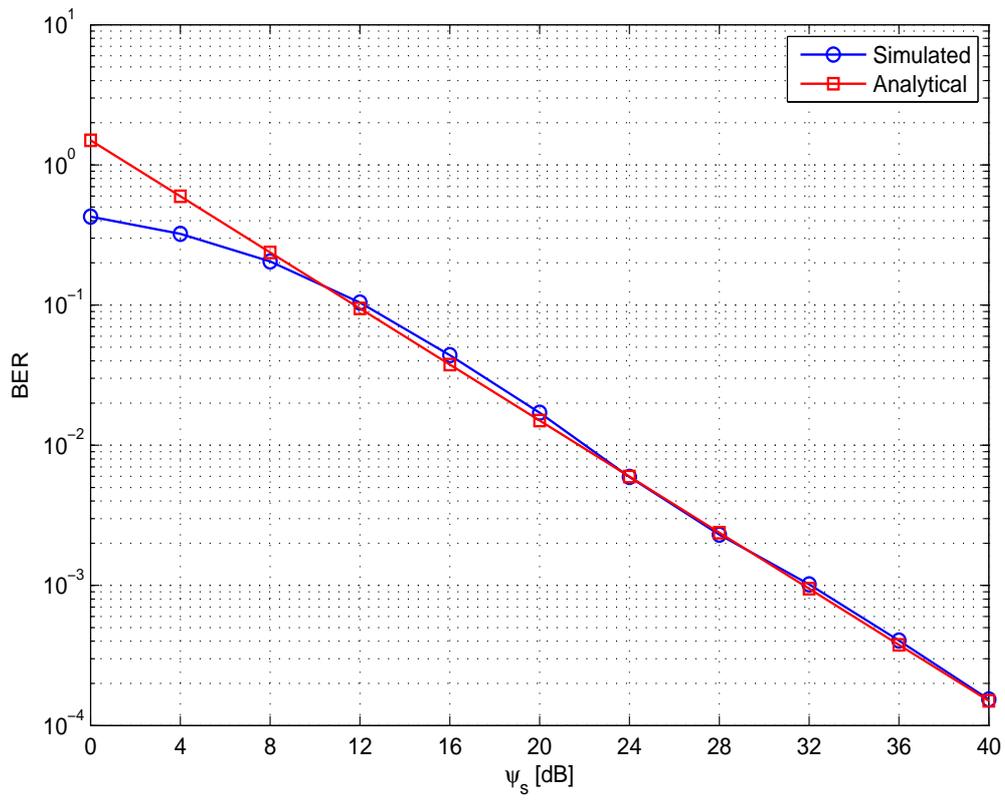}
\caption{Analytical and Simulated BER performance by the proposed
differential scheme, where $p_s=p_r=p/3$}
\label{Fig:Diff_bidirectional_relay_ana}
\end{figure}

In Fig.~\ref{Fig:Diff_bidirectional_relay_ana}, we compare the
analytical and simulated BER performance of the proposed
differential modulation scheme. Equal transmit power allocation is
also applied by setting $p_s=p_r=p/3$. From the figure, it can be
observed that at high SNR, the analytical BER derived by
(\ref{Eq:upPEPrs}) converged to the simulated result, which
justifies the validation of (\ref{Eq:upPEPrs}).

\begin{figure}[]
\centering
\includegraphics[height=4.5in,width=6.0in]{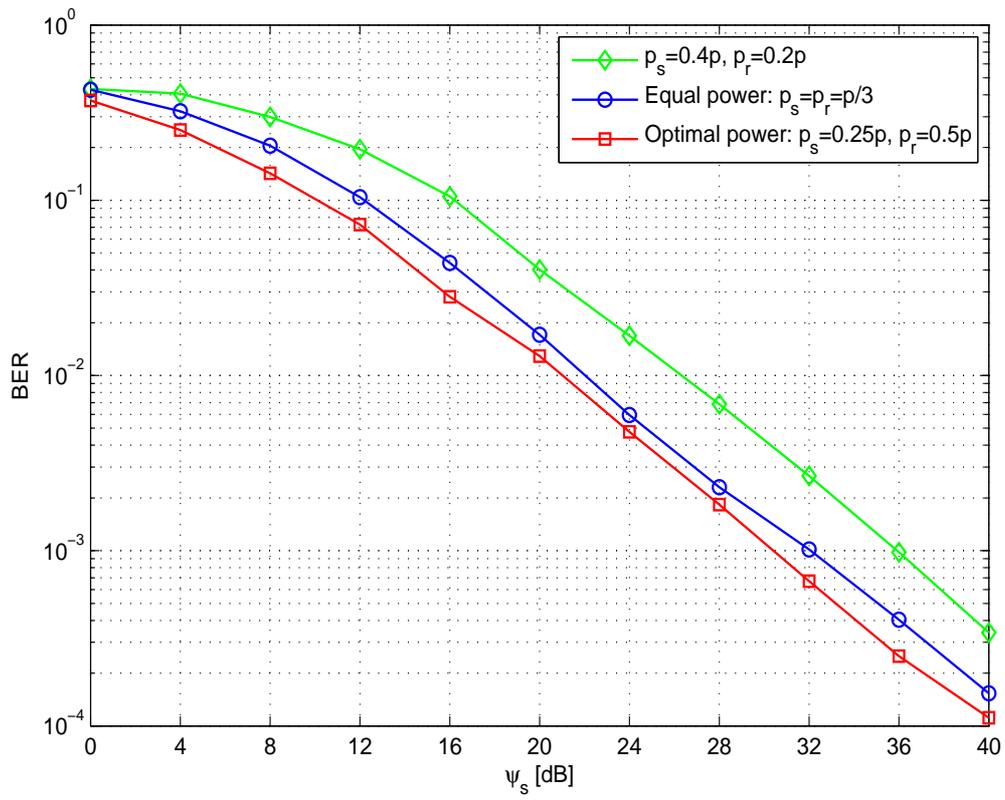}
\caption{Simulated BER performance by the proposed differential
scheme using transmit power allocation.} \label{Fig:Diff_bidirectional_relay_power}
\end{figure}

In Fig.~\ref{Fig:Diff_bidirectional_relay_power}, we examine the BER performance of the proposed differential modulation protocol with power allocation by setting $p_s=p/4$ and $p_r=p/2$ subject to the total power constraint. From Fig.~\ref{Fig:Diff_bidirectional_relay_power}, it can be observed that with optimal power allocation, the proposed scheme obtains about 2 dB performance gain in comparison with the equal power allocation scheme at high SNR. We also compare the result by another power setting: $p_s=0.4p$ and $p_r=0.2p$, and inferior result can be again observed compared to the optimal power allocation.

%In Fig.~\ref{Fig:Diff_bidirectional_relay_ser_upperbound_power}, we compare the BER curves using exact expression in (\ref{Eq:aveSER}) and the asymptotic bound in (\ref{Eq:upPEPrs}) when transmit power allocation is used. We can see from the figure, the optimal power allocation solutions by solving (\ref{Eq:aveSER}) and (\ref{Eq:upPEPrs}) is almost the same.

\begin{figure}[]
\centering
\includegraphics[height=4.5in,width=6.0in]{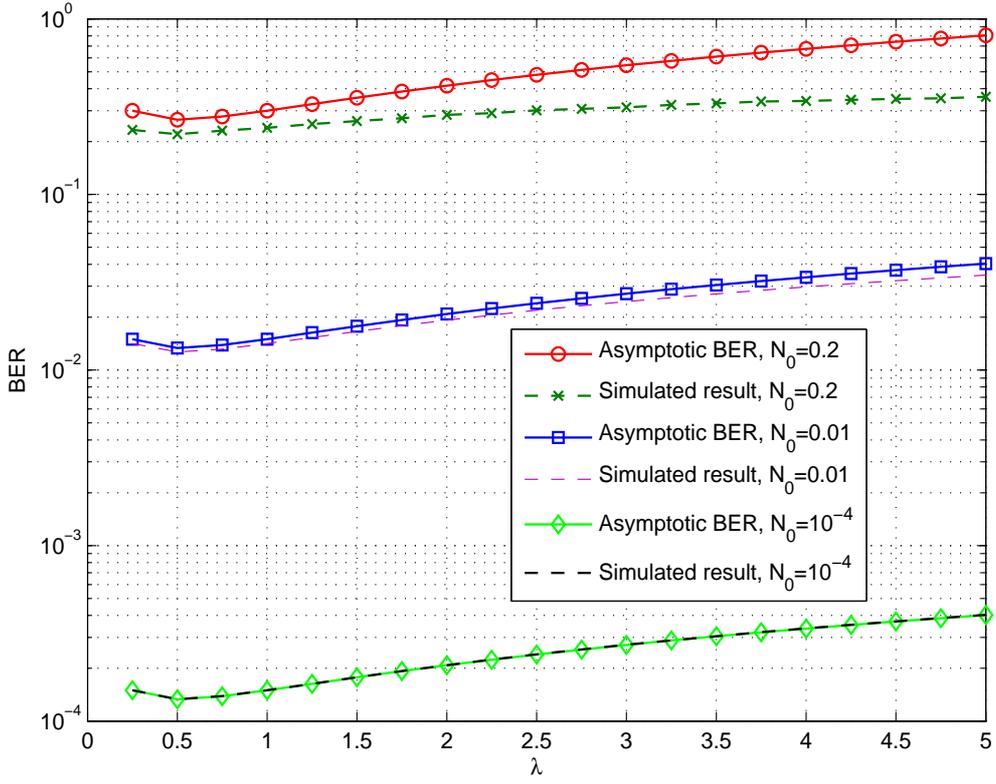}
\caption{Simulated BER performance by the proposed differential
scheme using transmit power allocation in term of $\lambda=p_s/p_r$,
with different noise variance $N_0$ and $\psi_s=\frac{\lambda{p}}{(2\lambda+1)N_0}$.}
\label{Fig:Diff_bidirectional_relay_ps_pr}
\end{figure}

In Fig.~\ref{Fig:Diff_bidirectional_relay_ps_pr}, we plot the BER curves in terms of $\lambda=p_s/p_r$ defined in (\ref{Eq:SNRy11}) using different noise variance $N_0$ with the asymptotic BER constraint in (\ref{Eq:upPEPrs}). With the power constraint $2p_s+p_r=p$, the SNR can be therefore calculated by $\psi_s=\frac{\lambda{p}}{(2\lambda+1)N_0}$, where $p=3$. It shows that best performance is obtained when $\lambda=0.5$. In other words, $p_s=p/4$ and $p_r=p/2$ is the optimal power setting between the sources and the relay, which further verify the power allocation strategy in (\ref{Eq:PowerAll}). From the figure, we can also see that the asymptotic BER is very close to the simulated results for various SNR values, and they result in the same power allocation solution of $p_s=p/4$ and $p_r=p/2$.

\begin{figure}[]
\centering
\includegraphics[height=4.5in,width=6.0in]{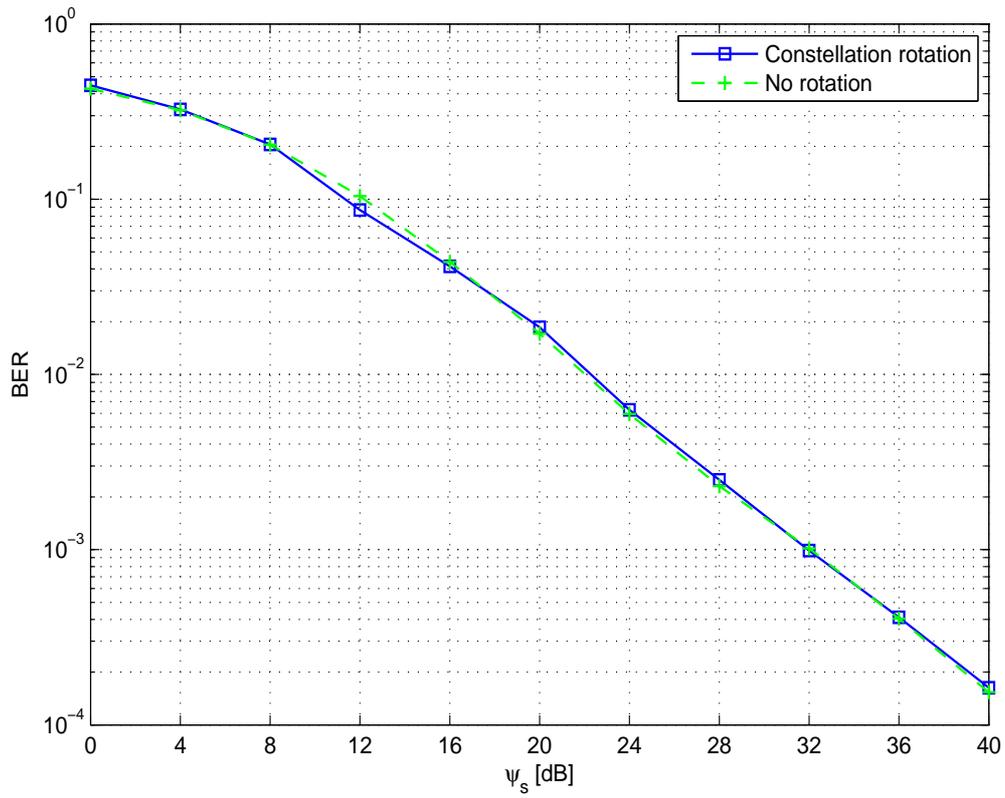}
\caption{Simulated BER performance by the proposed differential
scheme with and without using constellation rotation, where $p_s=p_r=p/3$.}
\label{Fig:Diff_bidirectional_relay_conste_rot}
\end{figure}

In Fig~\ref{Fig:Diff_bidirectional_relay_conste_rot}, we examine the
BER results of the proposed differential modulation scheme without
and with using constellation rotation, where the signal
constellation used by $S_1$ is rotated by $\pi/2$ relative to that
by $S_2$. It can be observed that the new result has very similar
with the curve without rotating constellations. This indicates that
using constellation rotation may not give system any gains given
large frame length.

\begin{figure}[]
\centering
\includegraphics[height=4.5in,width=6.0in]{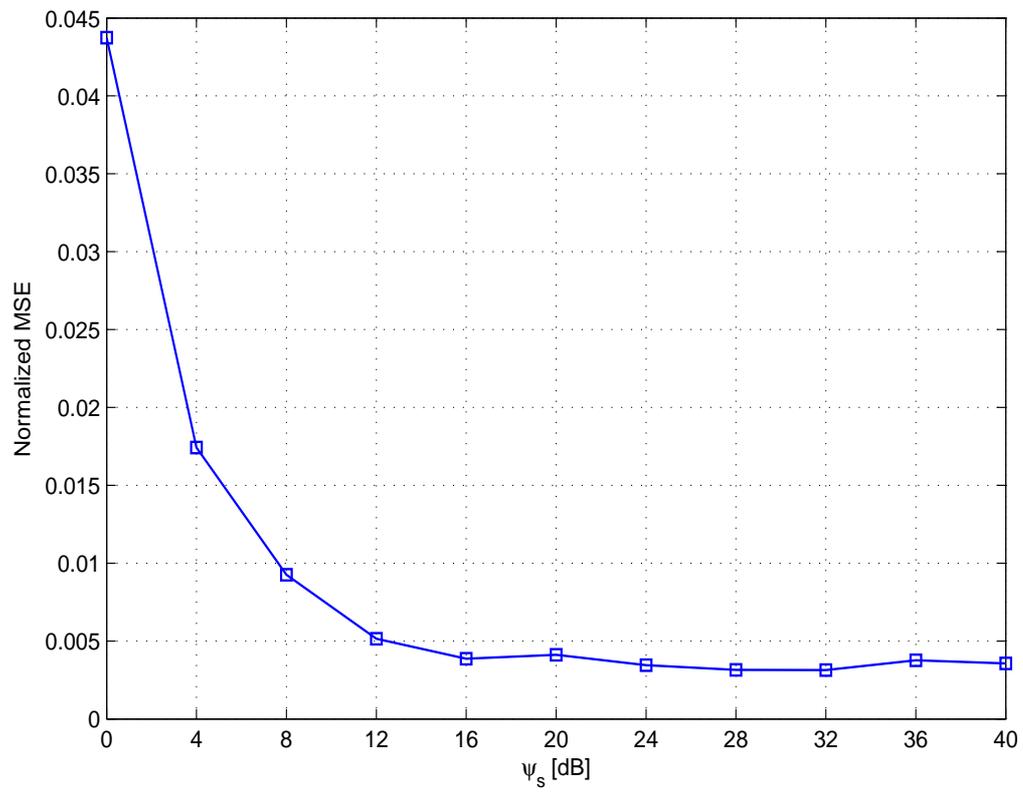}
\caption{The average normalized mean square error (MSE) of $\mu$ estimation in (\ref{Eq:muf}) as a function of the SNR.}
\label{Fig:standard_deviation_mu}
\end{figure}

Fig.~\ref{Fig:standard_deviation_mu} shows the average normalized mean square error (MSE) of $\mu$ estimation in (\ref{Eq:muf}) as a function of the SNR, where the average normalized MSE is calculated as $\frac{\frac{1}{N}\sum_{k=1}^N(\mu(k)-\widehat{\mu}(k))^2}{\mathbb{E}[\mu(k)]}$ and $\widehat{\mu}(k)$ is the estimate of $\mu(k)$. It can be observed that the estimation is quite accurate, particularly at high SNR.
%%%%%%%%%%%%%%%%%%%%%%%%%%%%%%%%%%%%%%%%%%%%%%%%%%%%%%%%%%%%%%%
\section{Conclusions}
%%%%%%%%%%%%%%%%%%%%%%%%%%%%%%%%%%%%%%%%%%%%%%%%%%%%%%%%%%%%%%%%%%%%%%%%%%%%%%%%%%%%%%%%%%%%%%%%%%%
In this paper, we have proposed a simple differential modulation
scheme for bi-directional relay communications using analog network
coding when neither sources nor the relay has access to channel
state information. Simulation results indicate that there exist
about 3~dB loss compared to the coherent detection scheme.
Analytical BER is derived to validate the proposed method. In
addition, based on the asymptotic BER at high SNR, an optimal power allocation between the sources and the relay was derived
to enhance the system performance.
%It is worthwhile pointing out that as $
%\gamma_d\approx{{\gamma_c}/{2}}$, the analytical SER for coherent
%detection with $M$-PSK constellations can be derived
%%%--
%\begin{align}
%    \text{SER}_c\approx\frac{(1+\lambda)(\psi_{s}^{-1}+\psi_{r}'^{-1})}{4g_\texttt{psk}}.
% \label{Eq:upPEPrsc}
%\end{align}
%%%--
%Hence, the same power allocation can be used by setting $p_s=p/4$
%and $p_r=p/2$.
%%%%%%%%%%%%%%%%%%%%%%%%%%%%%%%%%%%%%%%%%%%%%%%%%%%%%%%%%%%%%%%%%%%%%%%%%%%%%%%%%%%%%%%%%%%%%%%%%%%
%%%%%%%%%%%%%%%%%%%%%%%%%%%%%%%%%%%%%%%%%%%%%%%%%%%%%%%%%%%%%%%%%%%%%%%%%%%%%%%%%%%%%%%%%%%%%%%%%%%
%%%%%%%%%%%%%%%%%%%%%%%%%%%%%%%%%%%%%%%%%%%%%%%%%%%%%%%%%%%%%%%%%%%%%%%%%%%%%%%%%%%%%%%%%%%%%%%%%%
%\appendices

%%%%%%%%%%%%%%%%%%%%%%%%%%%%%%%%%%%%%%%%%%%%%%%%%%%%%%%%%%%%%%%%%%%%%%%%%%%%%%%%%%%%%%%%%%%%%%%%%%%
%%%%%%%%%%%%%%%%%%%%%%%%%%%%%%%%%%%%%%%%%%%%%%%%%%%%%%%%%%%%%%%%%%%%%%%%%%%%%%%%%%%%%%%%%%%%%%%%%%%

\pagebreak
%%---------------------------------------------------

\clearpage
%%---------------------------------------------------
%%----------------------------------------------------
\clearpage
%%---------------------------------------------------
%%----------------------------------------------------
\clearpage
%%---------------------------------------------------
%%----------------------------------------------------
\clearpage
%%%---------------------------------------------------
%\begin{figure}[]
%\centering
%\includegraphics[height=4.5in,width=6.0in]{Diff_bidirectional_relay_ser_upperbound_power.eps}
%\caption{Simulation results of the proposed differential
%scheme using transmit power allocation based on the exact BER and the asymptotic BER, where $p_s=p/4$ and
%$p_r=p/2$.} \label{Fig:Diff_bidirectional_relay_ser_upperbound_power}
%\end{figure}
%%%----------------------------------------------------
%\clearpage
%%---------------------------------------------------
%%----------------------------------------------------
\clearpage
%%---------------------------------------------------
%%----------------------------------------------------
\clearpage
%%---------------------------------------------------
%%----------------------------------------------------
\end{document}